# High $T_c$ superconductivity in a critical range of micro-strain and charge density in diborides

S. Agrestini, D. Di Castro, M. Sansone, N. L. Saini,
*Research Unit of INFM, Università di Roma "La Sapienza", P. le Aldo Moro 2, 00185 Roma, Italy*

A. Saccone, S. De Negri, M. Giovannini
*Dipartimento di Chimica e Chimica Industriale, Università di Genova, Via Dodecaneso 31, 16146 Genova, Italy*

M. Colapietro
*Dipartimento di Chimica, Università di Roma "La Sapienza", P. le Aldo Moro 2, 00185 Roma, Italy*

A. Bianconi
*Dipartimento di Fisica, Università di Roma "La Sapienza", P. le Aldo Moro 2, 00185 Roma, Italy*

(31 July 2001)

Expansion of the superlattice of boron layers, with $AB_2$ structure, due to different intercalated A atoms has been studied to understand the emergence of high $T_c$ superconductivity in the diborides. The structure of these metal heterostructures at the atomic limit (MEHALs) (with A=Al, Mg, Ti, Hf, Zr) has been measured by synchrotron x-ray diffraction. The increasing atomic radius of the intercalated A ions induces an increase of 1) the separation between the boron layers and 2) the tensile micro-strain $\varepsilon$ of the B-B distance within the boron layers. The results show that the superconductivity in these MEHAL appears in a critical region in a phase diagram controlled by two variables, the micro-strain and the charge density ($\varepsilon$, $\rho$).

PACS Codes: **74.70.Ad, 74.62.Dh; 74.80.Dm**

$T_c$ amplification in superconductors made of metal heterostructures at the atomic limit [MEHAL] was found in doped cuprate perovskites [1]. In the $Bi_2Sr_2CaCu_2O_{8+\delta}$ system the one-dimensional ordering of both the intercalated oxygen ions between the $CuO_2$ planes and the polaron ordering in the $CuO_2$ plane produces a heterogeneous metal made of a superlattice of quantum stripes. The $T_c$ amplification occurs at the optimum doping by tuning the chemical potential at the "shape resonance" of the superlattice. This resonance occurs near the dimensional 1D-2D cross-over of the topology of the Fermi surface [2]. At beginning of the 2001 Jun Akimitsu reported evidence of superconductivity at 39K in a superlattice of quantum wells, boron layers intercalated by Mg ions, a system known as *MgB₂* [3]. This system shows the superconducting phase [4] with a London penetration depth $\lambda \sim 140$ nm, a short Pippard coherence length $\xi_0 \sim 5.2$ nm, and an isotope coefficient $\alpha \sim 0.25$ [5]. It has been shown [6,7] that the high $T_c$ is obtained by tuning the chemical potential at the "shape resonance" of the superlattice of quantum wells. This is characterized by the dimensional 2D-3D cross-over of the topology of the Fermi surface, consistent with the patented process [8] of increasing the critical temperature in metal heterostructures at the atomic limit (MEHAL). According with this process the "shape resonance" can be reached by changing both the charge density and the structure of the superlattice. The fact that the critical temperature increases by increasing the separation between the metal units (and therefore the period of the superlattice) due to intercalation of larger ions, predicted in [2], has been recently verified experimentally in the superlattice of quantum dots: the metallic empty carbon spheres ($C_{60}$ buckyballs) separated by tribromomethane that reaches $T_c=117K$ [9].

In this work we have studied the variation of the separation between the boron metal layers as a function of the atomic radius of the intercalated ion. In fact the superlattice expands with increasing the atomic radius of intercalated ions. This structural parameter is relevant since it controls the band dispersion along the c-axis direction and therefore it controls $T_c$ via both the 2D-3D cross-over of the Fermi surface and the induced variation of the density of states.

The second key parameter that controls the high $T_c$ superconductivity in MEHALs is the elastic micro-strain in the metallic units modulated by the lattice mismatch with the intercalated materials [10]. This structural parameter increases the lattice anharmonicity and modulates the electron lattice interaction and therefore the pairing strength in the metallic units. Therefore we have investigated the tensile micro-strain in the boron layers induced by increasing the atomic radius of the intercalated ions. We show that diborides showing high $T_c$ occupy a critical region in a phase diagram with two variables, charge density and B-B micro-strain in the metallic B-monolayers.

The light metal diborides $AB_2$ (A=Mg, Al) were discovered as residuals in the chemical processing for the reduction of boron oxide with electropositive metals to obtain elemental boron [11] at the beginning of the XX century. The synthesis and characterization of aluminum diboride ($AlB_2$) [12] was reported in 1935 and magnesium diboride ($MgB_2$) in the 1950's [13-14]. With the development of the nuclear power industry, the borides (and boron carbide) were used for control rods and neutron shields and the actinide diborides (A=U, Pu) were studied. Recently the highly conducting and extremely hard transition metal diborides A=Ti, Zr, Cr have been used for turbine blades, combustion chamber liners, rocket nozzles, and ablation shields [15-19].

The diborides (e.g. $MgB_2$), having omega phase structure, show a natural nano-architecture made by two dimensional (2D) metallic boron (B) monolayers (in the xy plane), intercalated with epitaxial (A) monolayers forming an ABABAB heterostructure, a superlattice with a period determined by the *c* axis. The boron layers are made of a graphite-like net with the honeycomb lattice shown in figure 1. Short B-B distances form the metallic boron layers with covalent B-B bond. In the epitaxial hexagonal A layer, the atoms (A=Mg) are at the hollow site of the boron graphite-like layers and each A atom is coordinated by 12 boron atoms with an ionic bond, with the A atoms donating the valence electrons to the boron conduction band. The electronic band structure has been first calculated in the 1970s [20] and recently by advanced band structure calculations [21, 22].

We have studied commercially available $AB_2$ intermetallic compounds with transition metal atoms A of the Group IVB (A=Ti, Zr, Hf) from Alfa-Aesar with superconducting temperature $T_c$<0.3K, and they are compared with $AlMgB_4$, $Al_2MgB_6$ and $MgB_2$ samples. The $AlMgB_4$, $Al_2MgB_6$ and $MgB_2$ samples were synthesized by direct reaction of the starting materials of elemental magnesium and aluminum (rod, 99.9 mass% nominal purity) and boron (99.5 % pure <60 mesh powder). The elements in a stoichiometric ratio were enclosed in tantalum crucibles sealed by arc welding under argon atmosphere. The Ta crucibles were then sealed in heavy iron cylinder and heated for one hour at 800 °C and two hours a 950 °C in a furnace. The superconducting properties were investigated by the temperature dependence of the complex conductivity using the single-coil inductance method [6] showing $T_c$=4±2 K, 24±3 K, 39±0.5 K in $AlMgB_4$, $Al_2MgB_6$ and $MgB_2$ respectively.

The structure of the samples has been studied by x-ray diffraction using synchrotron radiation. We have used the wiggler source, at the third generation Elettra storage ring (Trieste), operated at 2 Gev and 170 mA. The diffraction patterns were recorded on the XRD beam line using synchrotron light and a double crystal Si (111) monochromator and it was focused on the sample by a Pt coated silicon mirror. A CCD detector of diameter 165 mm with 2048x2048 pixels per frame, from Mar-research, was used to record the diffraction patterns.

Figure 2(a) shows the x-ray powder diffraction (XRD) patterns of $TiB_2$, $ZrB_2$, $HfB_2$ while the figure 2(b) shows the patterns of $AlMgB_4$, $AlMg_2B_6$ and $MgB_2$, measured at low temperature T=100K. The $MgB_2$ samples show a minority phase of metallic Mg that does not appear in $AlMgB_4$ and $AlMg_2B_6$. These two last samples show broad peaks due to the small strained crystals and some additional superlattice reflections due to Al/Mg distribution.

The lattice parameters were determined by standard least-squares refinement of the diffraction data, using the GSAS program. The boron–boron distance $R_{BB}$ and the c/a ratio as a function of the atomic radius of the A atom [23, 24] are plotted in figure 3. The $AlMgB_4$ and $AlMg_2B_6$ phases are denoted by their stochiometry, 114 and 126 respectively.

An ideal omega phase ($AB_2$) is stable in the range of boron-boron distances 170 pm<$R_{BB}$<182 pm for 1<c/a<1.15. For the B-B distance and the c-axis larger than a critical value the lattice becomes a rumpled omega phase as known for $ReB_2$ where each boron atom is 182 pm distant from three others and the c-axis is bigger than the critical value. The boron atoms form puckered layers as in $ReB_2$ which shows a structure similar to the hexagonal case except that the two B-atoms of the graphite net are displaced slightly out of the (001) plane alternatively up and down in the direction of the *c* axis. For small A atomic radii as in $VB_2$, the $AB_2$ structure is not stable. In fact the XRD of the $VB_2$ sample from Alfa-Aesar shows the orthorhombic Cmcm 63 space group of uranium in agreement with electron diffraction data reported on EMS–on-line [25]. Moreover $BeB_2$ is unstable and shows both the $AlB_2$ structure [26] and the hexagonal $BeB_3$ structure [27] depending on the preparation conditions.

In figure 4 we report the ratio *c/a* versus the B-B distance. In the diborides there is no lattice mismatch between the boron layers and A layers for the *c/a* ratio 1.07457. For compounds with *c/a* >1.07457 the boron layer is under tensile stress with the micro-strain of the B-B distance ($R_{BB}$) and there is a compressive stress in the hexagonal A layer. Increasing the atomic radii of the spacer atom A in the same Group (e.g., for the Group containing Ti, Zr, Hf) of elements, the *c/a* ratio increases. The diborides show stable $AB_2$ omega phase for Al and Mg while for the larger micro-strain associated with larger atoms (such as Ca) this phase is no longer stable. As a matter of fact, we were not able to introduce even a few percent of Ca in the $AB_2$ lattice.

Increasing the atomic radii of the spacer in the $AB_2$ lattice changes not only the separation between the boron layers but also the B–B distance within the boron layers. We use as a measure of this lattice mismatch the micro-strain $\varepsilon=(a_0-a)/a_0=1-1.07457a/c$, since the expected value for the *a* axis of an ideal unstrained material is $a_0=c/1.07457$.

We could estimate the charge number density in the conduction bands, given by 3 electrons per boron ion plus n electrons donated by the spacer (e.g. n=2 for Mg), resulting an electron number density $\rho= (6+n)/(\sqrt{3}/2)a^2$ (i.e., in the range of 0.9 – 1 Å$^{-2}$).

The B-B micro-strain as a function of surface charge density is shown in figure 5 for different diborides. The lines denoted by $E_c$ and $E_A$ indicate where the Fermi level is tuned to the 2D-3D Fermi surface cross-over [6] and at the top of the σ sub-band respectively. The "shape resonance" for $T_c$ amplification occurs in a region around the $E_c$ line. We observe that high $T_c$ occurs also in a defined range of micro-strain [1-2] between

0.02 and 0.06. The phase diagram in figure 5 recalls the recently reported phase diagram for the cuprates [10], where superconductivity appears near a critical region of charge density and micro-strain of the $CuO_2$ plane.

In summary we have studied the diborides to explore a possible origin of the newly discovered high $T_c$ superconductivity in $MgB_2$. We have prepared $MgB_2$ ($T_c$~39 K) and measured the lattice parameters by x-ray diffraction on a series of diborides. We find that the superconductivity appears near a region in a two variables (micro-strain and charge density) phase diagram at ($\varepsilon$, $\rho$) where the Fermi level is tuned to the superconducting "shape resonance" for the metallic superlattice of boron layers [1,2,6-8] and the micro-strain triggers the electron-lattice interaction in a critical region. This provides a road-map for further investigation of superconductivity at high $T_c$ since using different approach (pressure, chemical doping, etc.) it is possible to move from the point of the highest $T_c$ in different directions.

**Acknowledgement**

This research has been supported by i) Progetto 5% Superconduttività del Consiglio Nazionale delle Ricerche (CNR); ii) Istituto Nazionale di Fisica della Materia (INFM); and iii) the Ministero dell'Università e della Ricerca Scientifica (MURST).

**Figure Captions**

Figure 1.  Pictorial view of the metallic boron monolayers (B) made of graphite-like honeycomb lattice separated by hexagonal Mg layers (A) forming a superlattice ABAB in the *c*-axis direction of metallic (B) layers intercalated by the layers (A).

Figure 2.  The X-ray diffraction pattern of several diborides measured with synchrotron radiation: a) $TiB_2$, $HfB_2$ and $ZrB_2$; b) $MgAlB_4$, $AlMg_2B_6$ and $MgB_2$.

Figure 3.  The *c/a* ratio and the B-B distance versus the atomic radii of atom A in the $AB_2$ compounds. Superconducting material $AlMgB_4$, is indicated as (114) $AlMg_2B_6$ is indicated as (126).

Figure 4.  The *c/a* ratio versus the B-B distance. There is no lattice mismatch and no micro-strain for the ratio *c/a*=1.07457. The dashed circle indicates the region of superconducting materials $AlMgB_4$, $AlMg_2B_6$ and $MgB_2$. The data for $BeB_2$ are taken from ref. 26.

Figure 5.  The phase diagram of diborides as a function of the surface charge number density in the boron planes and the micro-strain $\varepsilon=1-1.07457a/c$ of the boron layer. We have indicated the dashed region of superconducting materials $AlMgB_4$, with $T_c=4\pm2$ K; $AlMg_2B_6$, with $T_c=24\pm3$ K, and $MgB_2$ with $T_c=39\pm0.5$ K. High $T_c$ superconductivity appears in a critical region characterized by a critical range of micro-strain and charge density. The line $E_A$ indicates the location of the top of the boron $\sigma$ band [1-2] and the line $E_c$ indicates the critical point for the 2D to 3D crossover of the Fermi surface due to the boron $\sigma$ band [6-8].

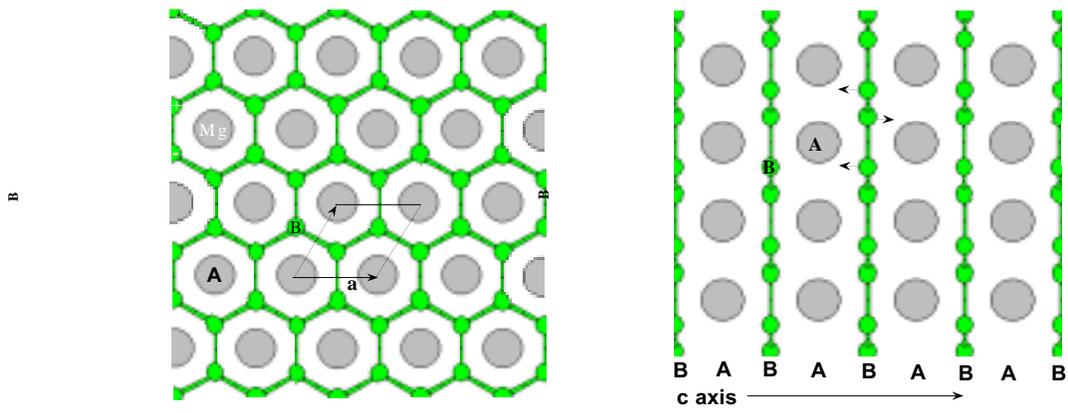

Fig. 1

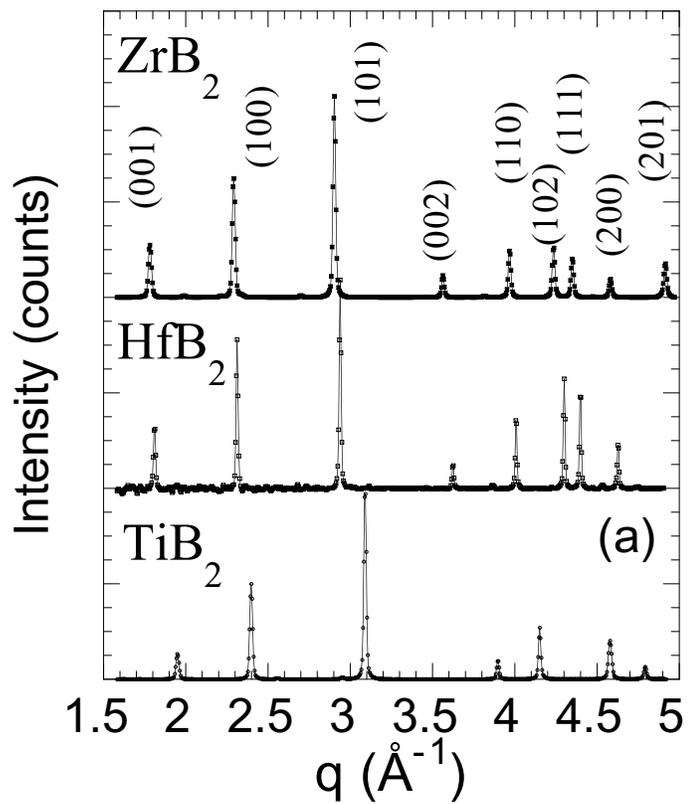

Fig. 2(a)

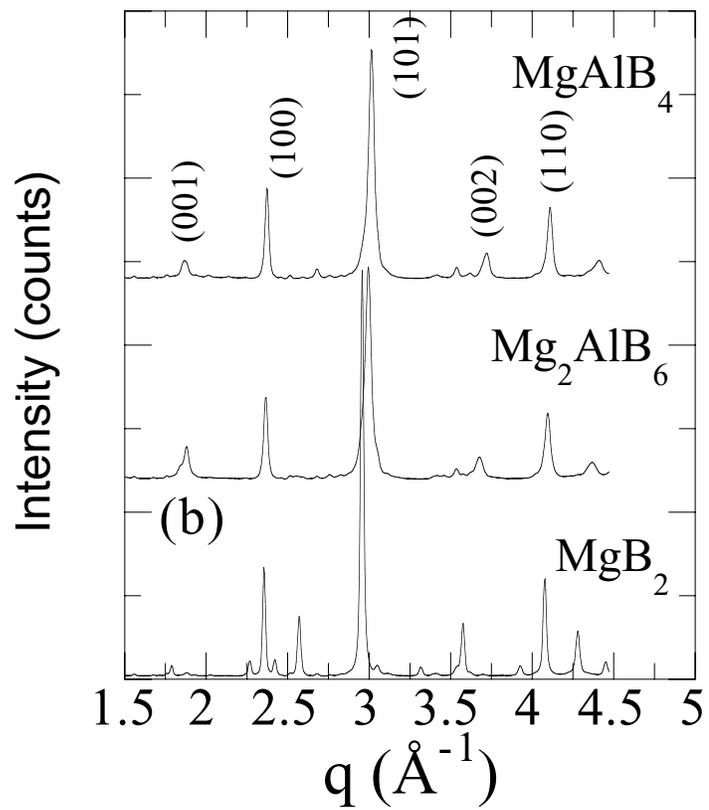

Fig. 2(b)

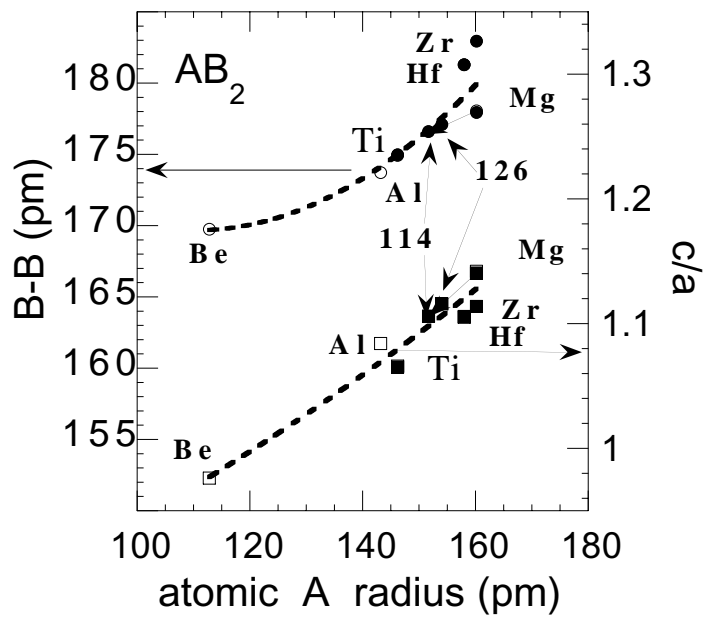

Fig. 3

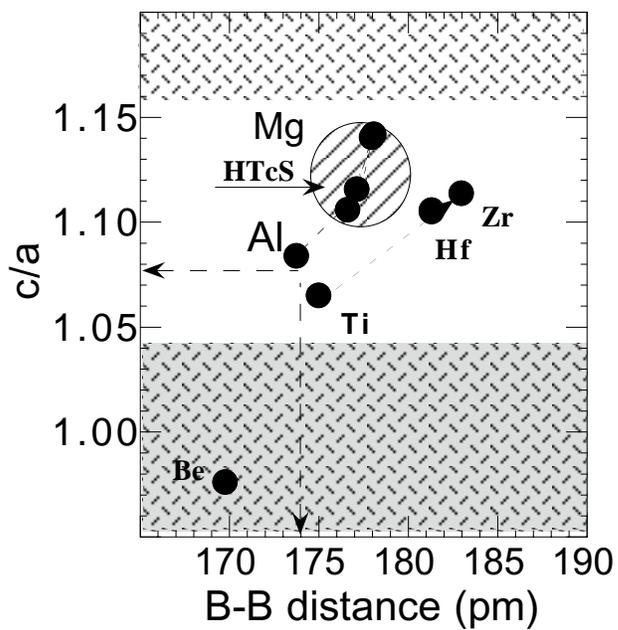

Fig. 4

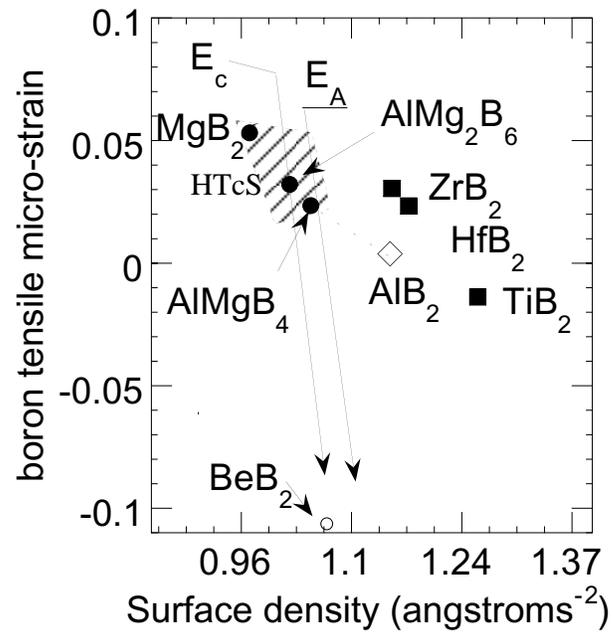

Fig. 5